\documentclass[a4paper]{PoS}

\title{Electromagnetic and transport properties of QGP within PLSM approach}

\ShortTitle{Electromagnetic and transport properties of QGP}

\author{\speaker{Abdel Nasser Tawfik}\\
        Egyptian Center for Theoretical Physics (ECTP), Modern University for Technology and Information (MTI), 11571 Cairo, Egypt, \\
World Laboratory for Cosmology And Particle Physics (WLCAPP), 11571 Cairo, Egypt\\
        E-mail: \email{a.tawfik@eng.mti.edu.eg}}

\abstract{In order to study the response of the quantum chromodynamic matter to finite electromagnetic fields, we utilize the Polyakov linear-sigma model (PLSM) in mean-field approximation. Due to participants' momentum imbalance and off-center relativistic motion of the spectators' electric charges, localized, short-lived, huge electromagnetic fields are to be generated in the relativistic heavy-ion collisions. We report on various electromagnetic and transport properties of the new-state-of matter; the quark-gluon plasma (QGP) within the QCD-like approach, PLSM. We find an excellent agreement between our PLSM calculations and various recent lattice QCD simulations and notice that the magnetization and the magnetic susceptibility and the relative permeability obviously increase in the QGP phase. We predict that increasing the magnetic field remarkable decreases the viscosity, especially in hadron phase. while in QGP phase, the viscous properties seem not being affected with. }

\FullConference{38th International Conference on High Energy Physics\\
		3-10 August 2016\\
		Chicago, USA}

\begin{document}

\section{Introduction}

One of RHIC's great discoveries was the unambiguous detection of signatures for the viscous properties of the new-state-of-matter; the quark-gluon plasma (QGP) \cite{Gyulassy:2004zy}. This experimental evidence agrees well with recent lattice QCD simulations that various thermodynamic quantities including viscosity are $15-20\%$ below the Boltzmann limit even at temperatures greater than the critical temperature, i.e. an indication for the formation of strongly correlated QGP matter (colored quarks and gluons)  \cite{Philipsen:2012nu}.

Recently, it was proposed that, huge, very localized electromagnetic fields are to be generated in the relativistic heavy-ion collisions. In additional to the off-center relativistic motion of the spectators' charges, the possible momentum imbalance of the participants contributes to the generation of short-lived electromagnetic fields, which likely have significant impacts on the QCD medium \cite{skovov}. Concrete experimental observables can be coupled to such gigantic magnetic fields ($\sim 10^{19}~$Gauss), i.e. at RHIC, $|q|B \simeq m_{\pi}^2$, while at LHC energies, $|q|B \simeq 10-15 m^2_{\pi}$, where $|q|$ is the net electric charge and $m_{\pi}$ is the pion mass \cite{Kharzeev:2015znc}. Furthermore, physics of early Universe, magnetors, acceleration of cosmic rays and creation of stars are additional examples, where such huge, very localized, short-lived electromagnetic fields are essential ingredients in improving our understanding of QCD matter under extreme conditions, such as temperatures, densities and now the electromagnetic stresses.

The present paper introduces a short summary on our recent calculations within the Polyakov linear-sigma model (PLSM) of the possible impacts of such electromagnetic fields on the QCD medium and on the transport properties of hadron and parton matter.

\section{Formalism}

Due to limited length of proceedings' pages, we omit details about the PLSM formalism. The readers are kindly advised to consult Refs. \cite{Tawfik:2016edq,Tawfik:2016ihn,Tawfik:2016gye,Tawfik:2016lih,Tawfik:2015apa,Ezzelarab:2015tya}. 
The magnetic catalysis and Landau quantization shall be implemented, where the magnetic field ($e B$) generated in the heavy-ion collisions is assumed to be aligned along the $z$-direction. At finite temperature ($T$), $f$-th quark's chemical potential ($\mu_f$), and $e B$, the PLSM antiquark-quark PLSM potential reads
\begin{eqnarray} 
\Omega_{  \bar{q}q}(T, \mu _f, B)&=& - 2 \sum_{f=l, s} \frac{|q_f| B \, T}{(2 \pi)^2} \,  \sum_{\nu = 0}^{\nu _{max_{f}}}  (2-\delta _{0 \nu })    \int_0^{\infty} dp_z \nonumber \\ && \hspace*{5.mm} 
\left\{ \ln \left[ 1+3\left(\phi+\phi^* e^{-\frac{E_{B, f} -\mu _f}{T}}\right)\; e^{-\frac{E_{B, f} -\mu _f}{T}} +e^{-3 \frac{E_{B, f} -\mu _f}{T}}\right] \right. \nonumber \\ 
&& \hspace*{3.9mm} \left.+\ln \left[ 1+3\left(\phi^*+\phi e^{-\frac{E_{B, f} +\mu _f}{T}}\right)\; e^{-\frac{E_{B, f} +\mu _f}{T}}+e^{-3 \frac{E_{B, f} +\mu _f}{T}}\right] \right\}, \label{PLSMqq}
\end{eqnarray}
where the factor $3$ assures that only states consisting of three quarks shall be taken into consideration, $|q_f|$ is the $f$-th quark's net electric charge and the dispersion relation becomes $E_{B, f}=[p_{z}^{2}+m_{f}^{2}+|q_{f}|(2n+1-\sigma) B]^{1/2}$.  $2-\delta_{0 \nu}$ stands for degenerate Landau Levels, where the maximum Landau level $\nu_{max_{f}} = \lfloor (\tau _f ^2 - \Lambda ^2 _{QCD})/(2 |q_f| B) \rfloor$. $2n+1-\sigma$ is to be replaced by a summation over Landau Levels, $0\, \leq \nu \, \leq \nu_{max_f}$, where $n$ stands of Landau level and $\sigma$ is related to the spin quantum number. Both Polyakov-loop and LSM mesonic potentials aren't affected by finite $e B$ and thus interested readers can recall Refs. \cite{Tawfik:2016edq,Tawfik:2016ihn,Tawfik:2016gye,Tawfik:2016lih,Tawfik:2015apa,Ezzelarab:2015tya} for details. 

At finite $e B$, the QCD matter responses through modifying the free energy density $\mathcal{F}(T, \mu _f, B)= - (T/V)  \ln \mathcal{Z}(T, \mu _f, B)$. Accordingly, the magnetization, for instance, can be deduced. Its sign determines whether QCD matter is {\it para}- or {\it dia}-magnetic, i.e. $\mathcal{M}>0$ or $\mathcal{M}<0$, respectively. Both magnetic susceptibility ($\chi_B$) and relative magnetic permeability ($\mu_r$) are strongly related to each other, 
\begin{eqnarray}
\mathcal{M} = - \frac{\partial \mathcal{F}(T, \mu _f, B)}{\partial (eB)}, \quad \chi_B = - \frac{\partial^2 \mathcal{F}(T, \mu _f, B)}{\partial (eB)^2}\vert_{eB=0}, \qquad 
\mu_r  = 1+\frac{1}{1 - 4 \pi \, \alpha_{f} \, \chi _B}, \label{permeability}
\end{eqnarray} 
where fine structure constant $\alpha_{f}=q_f^2/4\pi$ and $\mu_r$ depends on the induced magnetic filed ($B^{ind}$) normalized to the external one ($B^{ext}$); $\mu_r=1+B^{ind}/B^{ext}$. The temperature dependences of $\mathcal{M}$, $\chi_B$, and $\mu_r$ are illustrated in Fig. \ref{propes}.

In non-Abelian external magnetic field, various transport coefficients can be determined from relativistic kinetic theory. The relaxation time approximation can be applied to the Boltzmann-Uehling-Uhlenbeck (BUU) equations and by using Chapman-Enskog expansion, where bulk and shear viscosities can be expressed as \cite{Tawfik:2016ihn,Chakraborty:2011},
\begin{eqnarray}
\zeta (T, \mu_f, B) &=&  \frac{1}{9\, T}  \sum_f  \frac{|q_f|B}{2\pi} \sum_\nu \int \frac{dp}{2\pi} \left(2-\delta_{0\nu}\right) \, \frac{\tau_f}{E_{B,f} ^2} \, \left[\frac{|\vec{p}|^2}{3} - c_s^2 E_{B,f}^2 \right]^2 \, f_{f} (T, \mu_f, B) ,  \label{eq:buubulk} \\
\eta (T,\mu_f, B) &=&  \frac{1}{15\, T}  \sum_f  \frac{|q_f|B}{2\pi} \sum_\nu \int \frac{dp}{2\pi} \left(2-\delta_{0\nu}\right)\, \frac{p^4}{E_{B,f}^2}\,  \tau_f  f_f (T, \mu_f, B),  \label{eq:buushear}
\end{eqnarray} 
with $c_s^2$ being the speed of sound squared and the distribution function ($f_f$) is given as
\begin{eqnarray}
f_f(T, \mu_f, B) &=& \frac{\left(\phi^*+2 \phi \exp\left[-\frac{E_{B,f}-\mu _f}{T}\right]\right)\,  \exp\left[-\frac{E_{B,f}-\mu _f}{T}\right]+ \exp\left[-3 \frac{E_{B,f}-\mu _f}{T}\right]}{1+3\left(\phi^*+\phi  \exp\left[-\frac{E_{B,f}-\mu _f}{T}\right]\right)\,  \exp\left[-\frac{E_{B,f}-\mu _f}{T}\right]+ \exp\left[-3 \frac{E_{B,f}-\mu _f}{T}\right]}.
\label{fqaurk} 
\end{eqnarray}
The relaxation time ($\tau_f$) is modelled as $\tau_f \propto \left.T\right|_{T>T_c}$ and $\tau_f \propto \left.\exp(m_f/T)\right|_{T<T_c}$. The results on bulk and shear viscosity normalized to entropy are depicted in Fig. \ref{fig:Bulk_viscosity2}.

\section{Results and Conclusions}

\begin{figure}[htb]
\centering{
\includegraphics[width=3.cm,angle=-90]{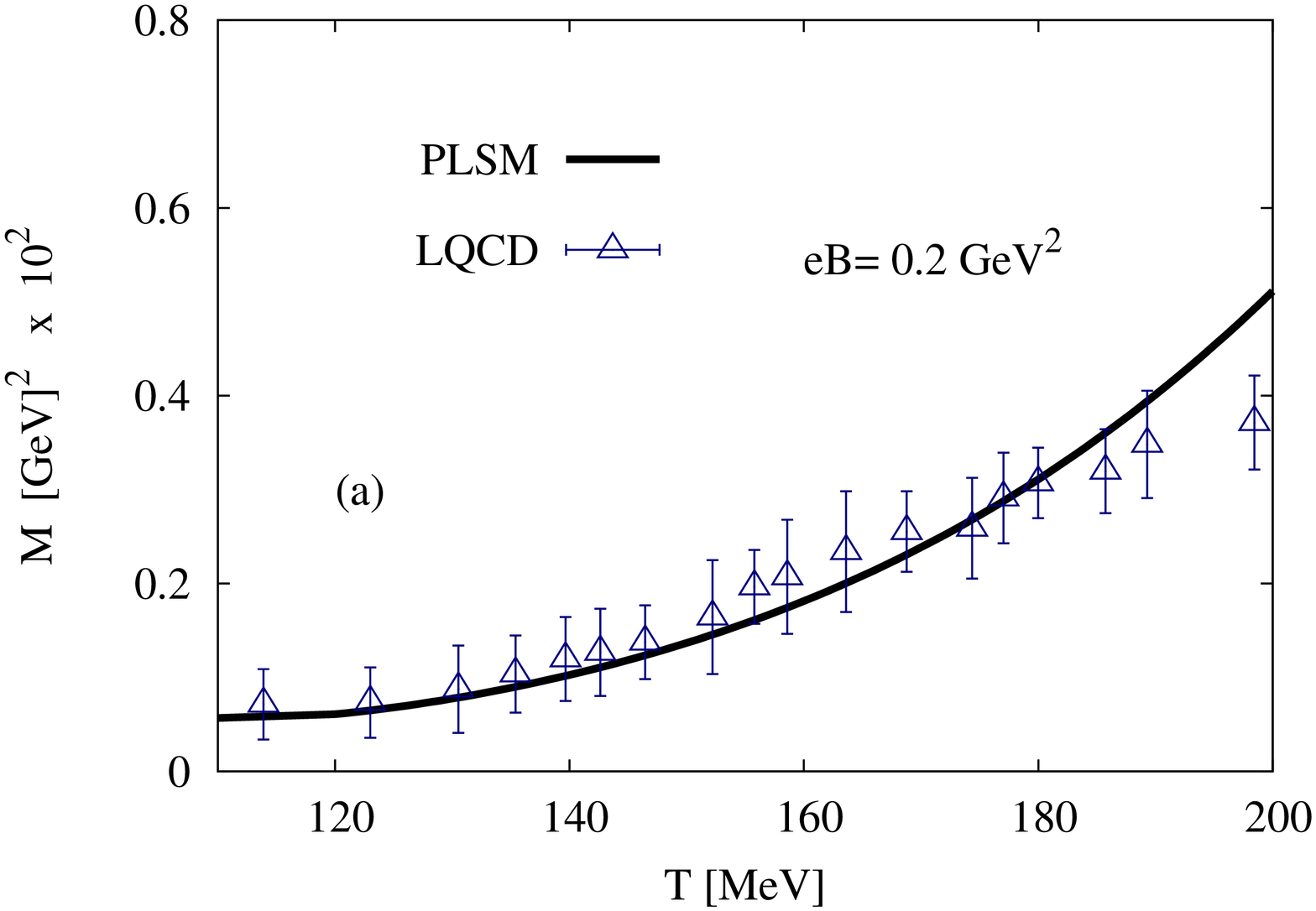}
\includegraphics[width=3.cm,angle=-90]{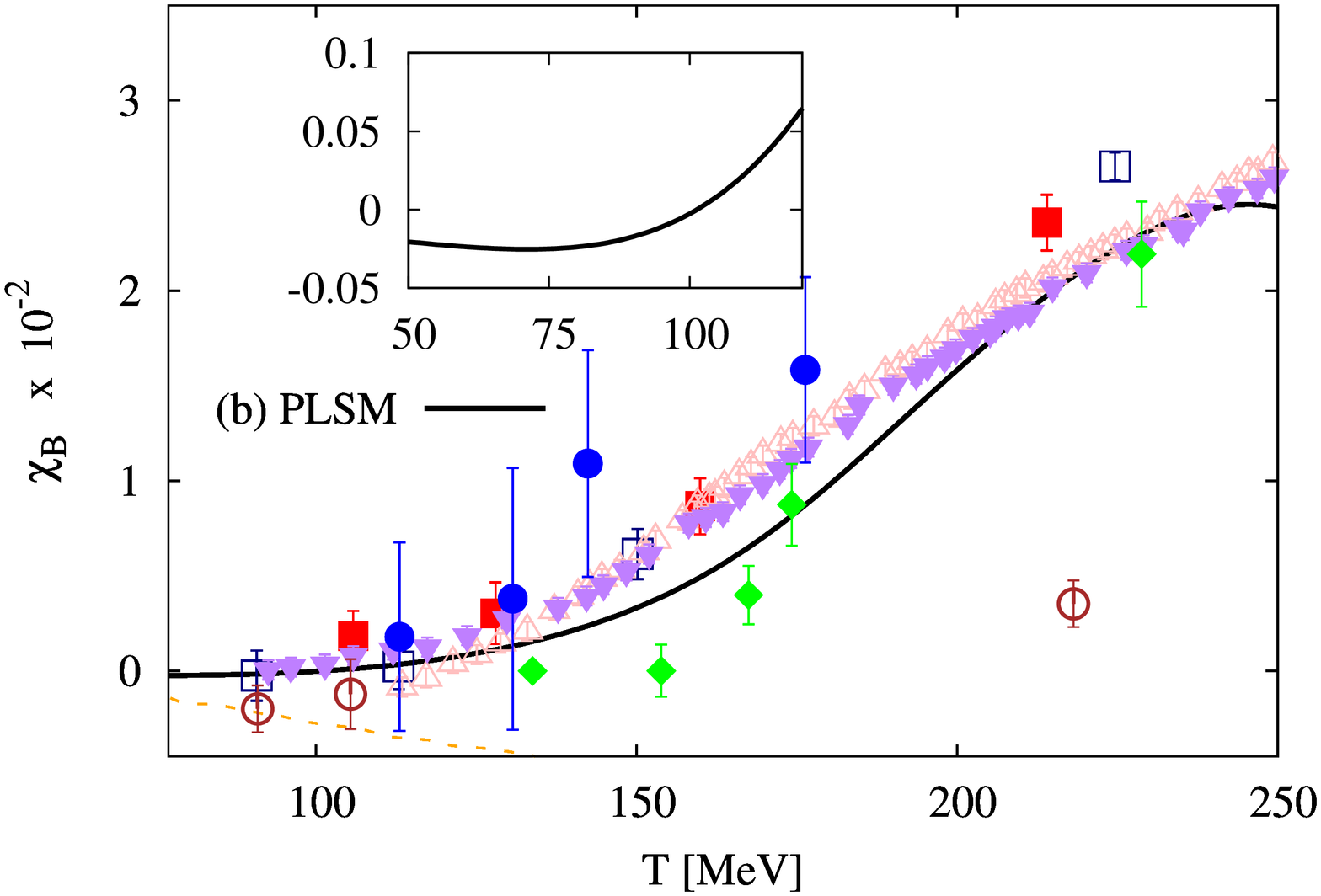}
\includegraphics[width=3.cm,angle=-90]{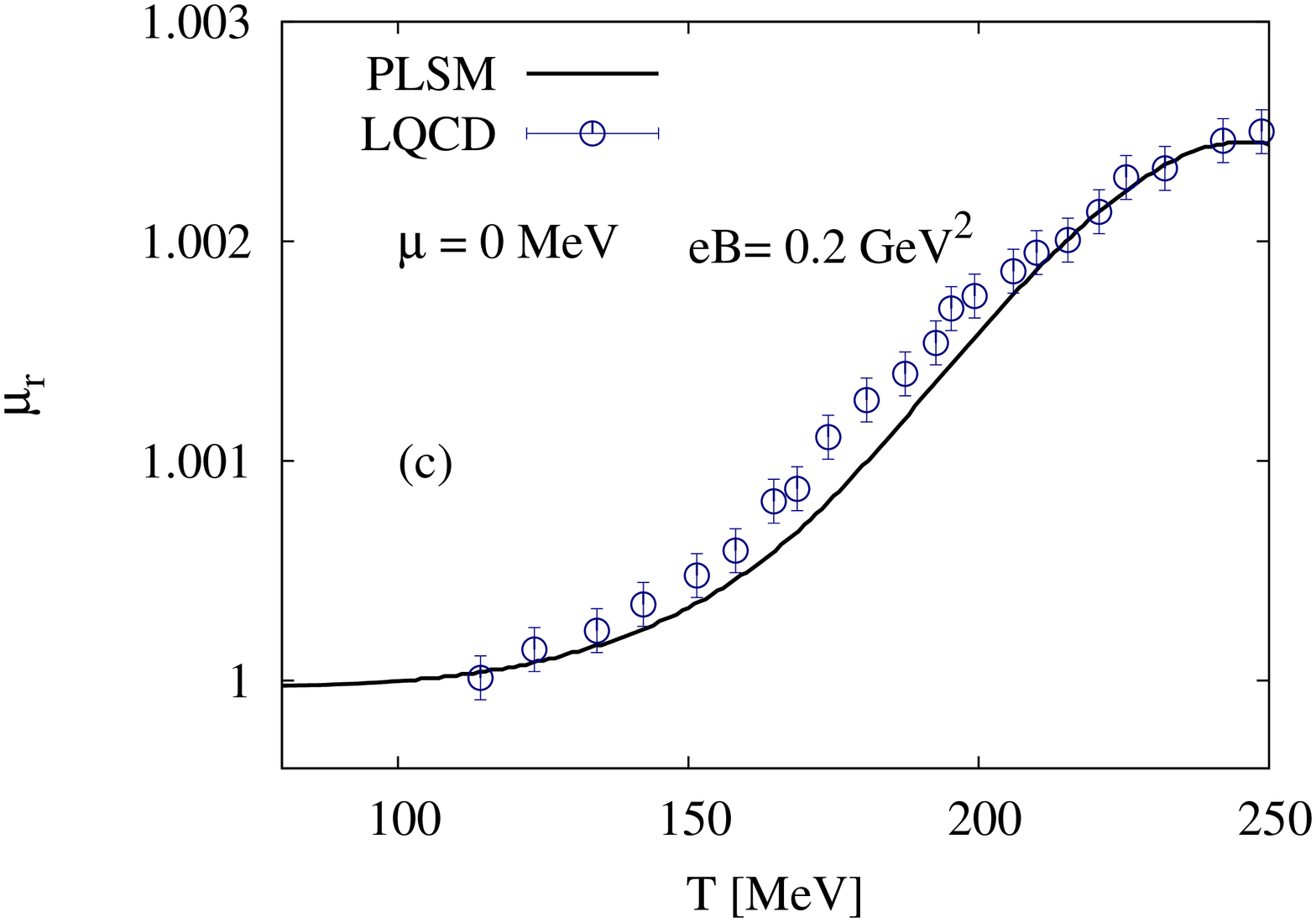}
\caption{\footnotesize At $\mu=0$, the temperature dependence of $\mathcal{M}$ (left-hand panel), $\chi_B$ (middle-panel) and $\mu_r$ (right-hand panel) are depicted. The results are confronted to different recent lattice calculations (symbols) \cite{lattice:2014}. \label{propes}
}}
\end{figure}

Figure \ref{propes} shows the temperature dependence of $\mathcal{M}$, $\chi_B$, and $\mu_r$ at $\mu=0$ and $e B=0.2~$GeV$^2$. We notice that $\mathcal{M}$, $\chi_B$, and $\mu_r$ apparently increase when the QCD matter  is converted from hadron to parton phase. It is obvious that, our PLSM calculations agree well with various recent lattice QCD simulations indicating that PLSM is excellently able to describe the transport properties of the QCD matter. On the other hand, PLSM can be utilized in predicting other properties, for instance, the results clearly show that the QGP at two times $T_c$ lay below the Boltzmann limit. At $T>T_c$, the resulting viscous properties approve such a conclusion.

\begin{figure}[htb]
\centering{
\includegraphics[width=4.cm,angle=-90]{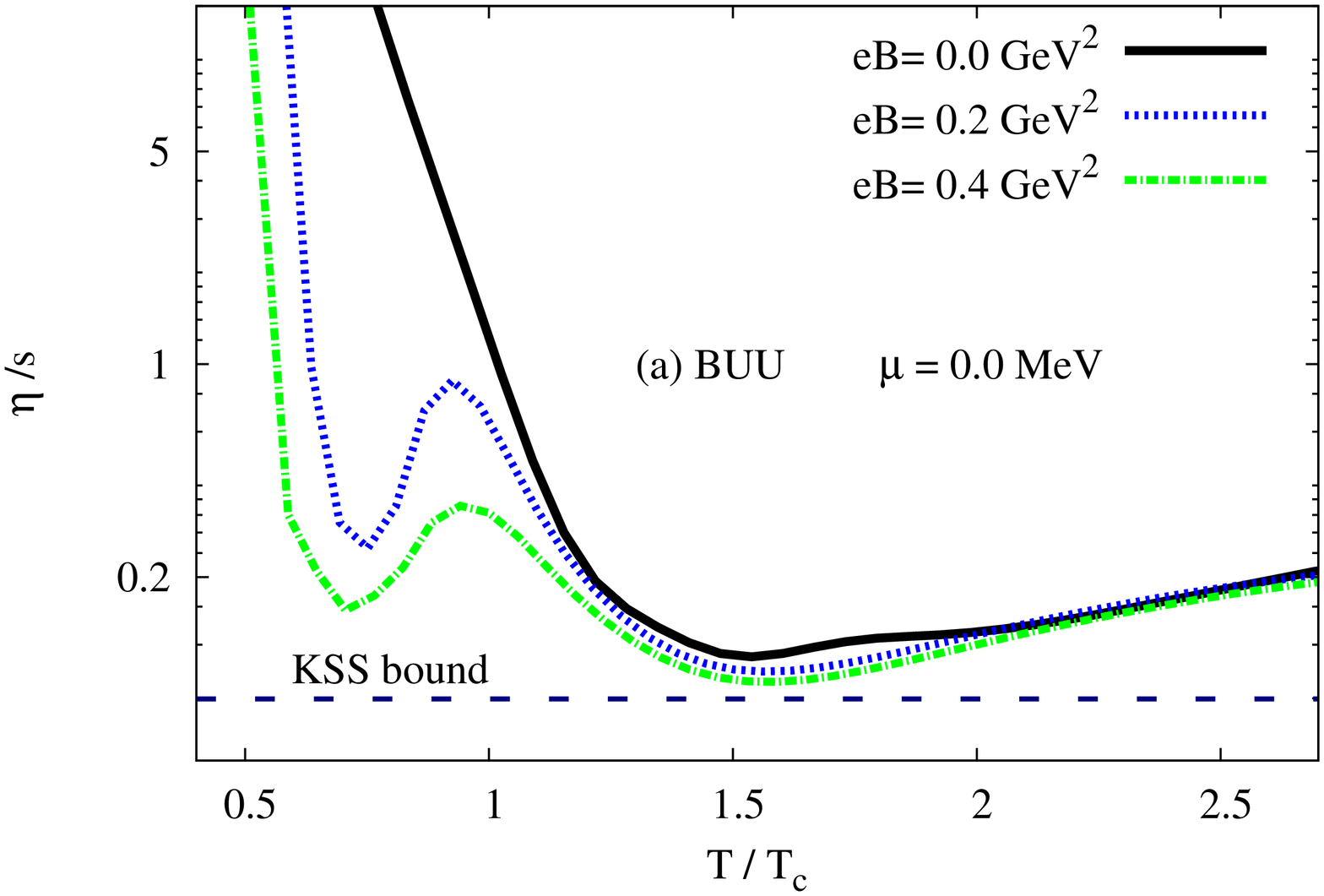}
\includegraphics[width=4.cm,angle=-90]{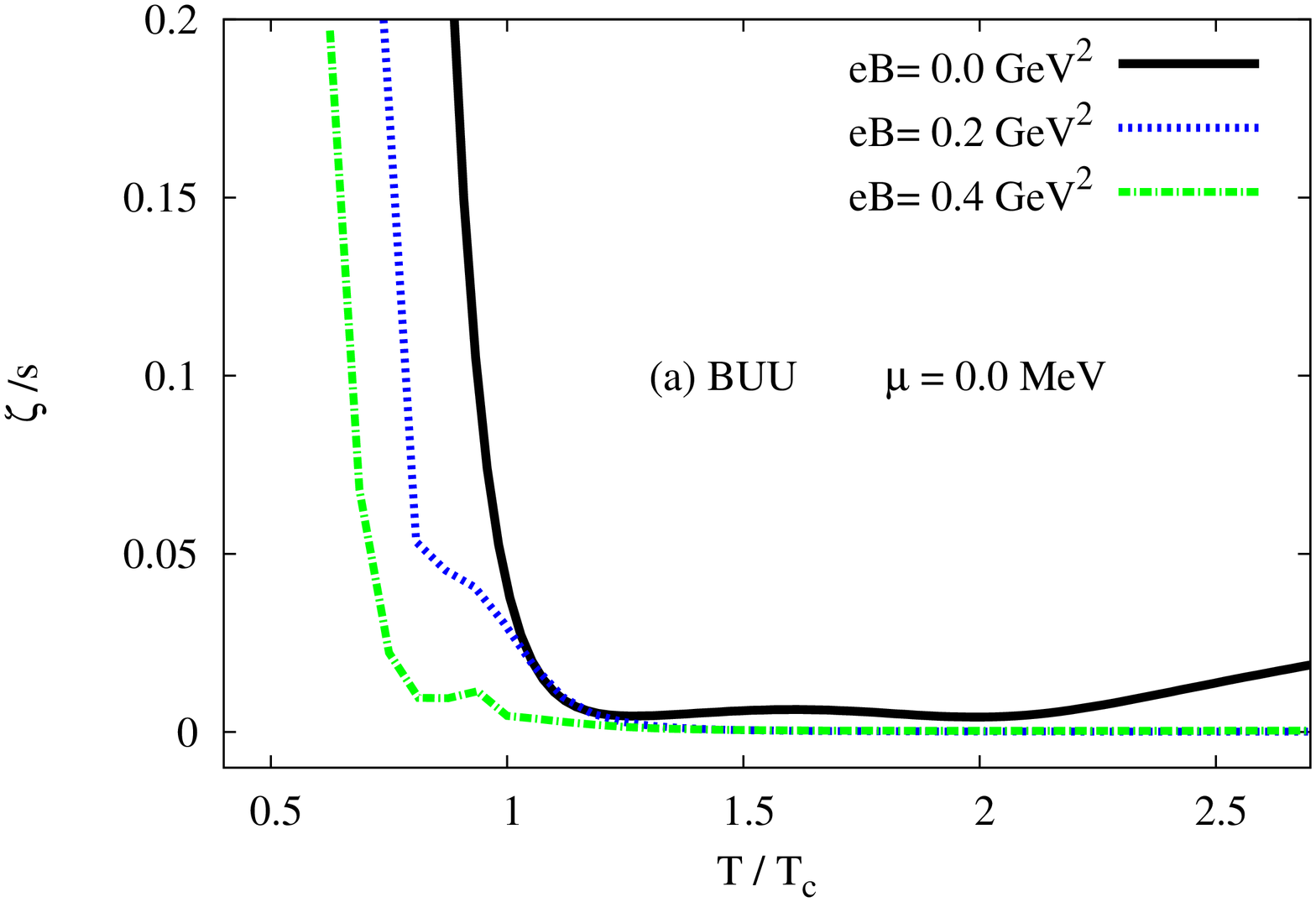}
\caption{\footnotesize Temperature dependences of $\eta/s$ (left-hand panel) and $\zeta/s$ (right-hand panel) are calculated from PLSM through BUU equations at $eB=0.0~$GeV$^2$ (solid), $eB=0.2~$GeV$^2$ (dotted) and $eB=0.4~$GeV$^2$ (dot-dashed curves) and $\mu=0$.  \label{fig:Bulk_viscosity2}
}}
\end{figure}

Figure \ref{fig:Bulk_viscosity2} presents the temperature dependences of $\xi/s$ (a) and $\eta/s$ (b) calculated from PLSM through BUU equations at $\mu=0$ and at $eB=0.0~$GeV$^2$ (solid), $eB=0.2~$GeV$^2$ (dotted) and $eB=0.4~$GeV$^2$ (dot-dashed curves). At $T>T_c$, the increase in $e B$ seems not affecting both $\xi/s$ and $\eta/s$, but at $T<T_c$, there is a remarkable decrease in $\xi/s$ and $\eta/s$ with increasing $e B$. The peaks at $T_c$ can be interpreted as the generated magnetic field  likely survives the critical temperature. Accordingly, this might affect the particle production and the deconfinement phase-transition. At $T>T_c$, the QGP viscosity doesn't only remain finite but slowly increases with the temperature.

\end{document}